\documentclass[sigconf]{acmart}
\usepackage{amsmath}
\usepackage{amsfonts}
\usepackage{bbold}
\usepackage{amsfonts}
\usepackage{booktabs} % for professional tables
\usepackage{multirow}
\usepackage{longtable}% for long tables
\usepackage{graphics}
\usepackage{subfigure}
\usepackage{bm}
\usepackage{bbm}
\usepackage[ruled]{algorithm2e}
\usepackage{setspace}
\usepackage{hyperref}
\hypersetup{
    colorlinks=true,
    linkcolor=blue,
    filecolor=blue,      
    urlcolor=blue,
    citecolor=blue,
}

\AtBeginDocument{%
  \providecommand\BibTeX{{%
    \normalfont B\kern-0.5em{\scshape i\kern-0.25em b}\kern-0.8em\TeX}}}

\copyrightyear{2022}
\acmYear{2022}
\setcopyright{acmcopyright}\acmConference[WWW '22]{Proceedings of the ACM Web Conference 2022}{April 25--29, 2022}{Virtual Event, Lyon, France}
\acmBooktitle{Proceedings of the ACM Web Conference 2022 (WWW '22), April 25--29, 2022, Virtual Event, Lyon, France}
\acmPrice{15.00}
\acmDOI{10.1145/3485447.3512208}
\acmISBN{978-1-4503-9096-5/22/04}

\acmSubmissionID{w09fp2707}

\begin{document}

\title{Graph Communal Contrastive Learning}

\author{Bolian Li}
\orcid{0000-0002-1977-0764}
\affiliation{%
  \institution{Tianjin University}
  \country{Tianjin, China}
}
\email{libolian@tju.edu.cn}
\author{Baoyu Jing}
\affiliation{%
  \institution{University of Illinois at\\Urbana-Champaign}
  \country{IL, USA}
}
\email{baoyuj2@illinois.edu}
\author{Hanghang Tong}
% \authornote{Corresponding author.}
\affiliation{%
  \institution{University of Illinois at\\Urbana-Champaign}
  \country{IL, USA}
}
\email{htong@illinois.edu}


\begin{abstract}
Graph representation learning is crucial for many real-world applications (e.g. social relation analysis). A fundamental problem for graph representation learning is how to effectively learn representations without human labeling, which is usually costly and time-consuming. Graph contrastive learning (GCL) addresses this problem by pulling the positive node pairs (or similar nodes) closer while pushing the negative node pairs (or dissimilar nodes) apart in the representation space. Despite the success of the existing GCL methods, they primarily sample node pairs based on the node-level proximity yet the community structures have rarely been taken into consideration. As a result, two nodes from the same community might be sampled as a negative pair. We argue that the community information should be considered to identify node pairs in the same communities, where the nodes insides are semantically similar. To address this issue, we propose a novel \textbf{G}raph \textbf{Co}mmunal C\textbf{o}ntrastive \textbf{L}earning ($gCooL$) framework to jointly learn the community partition and learn node representations in an end-to-end fashion. Specifically, the proposed $gCooL$ consists of two components: a Dense Community Aggregation ($DeCA$) algorithm for community detection and a Reweighted Self-supervised Cross-contrastive ($ReSC$) training scheme to utilize the community information. Additionally, the real-world graphs are complex and often consist of multiple views. In this paper, we demonstrate that the proposed $gCooL$ can also be naturally adapted to multiplex graphs. Finally, we comprehensively evaluate the proposed $gCooL$ on a variety of real-world graphs. The experimental results show that the $gCooL$ outperforms the state-of-the-art methods.

\end{abstract}




\maketitle

\section{Introduction}
Graph representation learning is crucial in many real-world applications, including predicting friendship in social relationships~\cite{zhu2017emotional}, 
recommending products to potential users~\cite{li2020personalized} and detecting social opinions~\cite{wang2017sentiment, DyDiff-VAE}. However, traditional graph representation learning methods~\cite{hu2019hierarchical,kipf2016semi,velivckovic2018graph} demand labeled graph data of high quality, while such data is too expensive to obtain and sometimes even unavailable due to privacy and fairness concerns~\cite{kang2021fair, kang2020inform}.
To address this problem, recent Graph Contrastive Learning (GCL) \cite{li2019graph,kipf2016semi,hamilton2017inductive,you2020graph,zhu2021graph,qiu2020gcc,hassani2020contrastive} proposes to train graph encoders by distinguishing positive and negative node pairs without using external labels. Among them, node-level GCL methods regard all node pairs as negative samples, which pulls closer positive node pairs (representations of the same node in two views) and pushes away negative node pairs (representations of different nodes) \cite{wu2021self}. The variants of GCL~\cite{sun2020infograph,hassani2020contrastive,zhang2020iterative} employ mutual information maximization to contrast node pairs and have achieved state-of-the-art performance on many downstream tasks, such as node classification~\cite{bhagat2011node}, link prediction~\cite{liben2007link} and node clustering~\cite{schaeffer2007graph}.

Despite the success of these methods, most of the existing GCL methods primarily focus on the node-level proximity but fail to explore the inherent community structures in the graphs when sampling node pairs. For example, MVGRL~\cite{hassani2020contrastive}, GCA~\cite{zhu2021graph} and GraphCL~\cite{you2020graph} regard the same nodes in different graph views as positive pairs; \cite{hassani2020contrastive} regards the neighbors as positive node pairs; DGI~\cite{velivckovic2018deep} and HDI~\cite{jing2021hdmi} produce negative pairs by randomly shuffling the node attributes; \cite{zhang2020motif,qiu2020gcc,hafidi2020graphcl} regard negative samples as different sub-graphs.
However, the structural information such as the community structures in the graphs has not been fully explored yet.
The community structures can be found in many real graphs~\cite{xu2018powerful}. For example, in social graphs, people are grouped by their interests~\cite{scott1988social}. The group of people with the same interest tend to be densely connected by their interactions, while the people with different interests are loosely connected. Therefore, the people in the same interest community are graphically similar and treating them as negative pairs will introduce graphic errors to the node representations.
To address this problem, we firstly propose a novel dense community aggregation ($DeCA$) algorithm to learn community partition based on structural information in the graphs. Next, we introduce a novel reweighted self-supervised contrastive ($ReSC$) training scheme to pull the nodes within the same community closer to each other in the representation space.

Additionally, in the real graphs, nodes can be connected by multiple relations \cite{park2020unsupervised, jing2021multiplex, jing2021network, yan2021dynamic}, and each relation can be treated as a graph view (known as multiplex graphs~\cite{de2013mathematical}). One way for graph representation learning methods to be adapted to the multiplex graphs is to learn the representations of each graph view separately and then combining them with a fusion model~\cite{jing2021hdmi}. However, this approach ignores the graphic dependence between different graph views. For example, IMDB~\cite{wang2019heterogeneous} can be modeled as a multiplex movie graph connected by both co-actor and co-director relations. These two kinds of connections are not independent since personal relations between actors and directors would affect their participation in movies. To address this problem, we apply contrastive training that is consistent with the situation of the single-relation graph, applying pair-wise contrastive training on each pair of graph views and using a post-ensemble fashion to combine model outputs of all views in the down-stream tasks.

The main contributions of this paper are summarized as follows:
\begin{itemize}
    \item We propose a novel Dense Community Aggregation ($DeCA$) algorithm to detect structurally related communities by end-to-end training.
    \item We propose a Reweighted Self-supervised Cross-contrastive ($ReSC$) training scheme, which employs community information to enhance the performance in down-stream tasks.
    \item We adapt our model to multiplex graphs by pair-wise contrastive training, considering the dependence between different graph views.
    \item We evaluate our model on various real-world datasets with multiple down-stream tasks and metrics. Our model achieves the state-of-the-art performance.
\end{itemize}
The rest of the paper is organized as: preliminaries (Sec.~\ref{sec:pre}), methodology (Sec.~\ref{sec:method}), experiments (Sec.~\ref{sec:exp}), related works (Sec.~\ref{sec:relate}) and conclusion (Sec.~\ref{sec:con}).

\begin{figure}[t!]
\centering
\includegraphics[width=8cm]{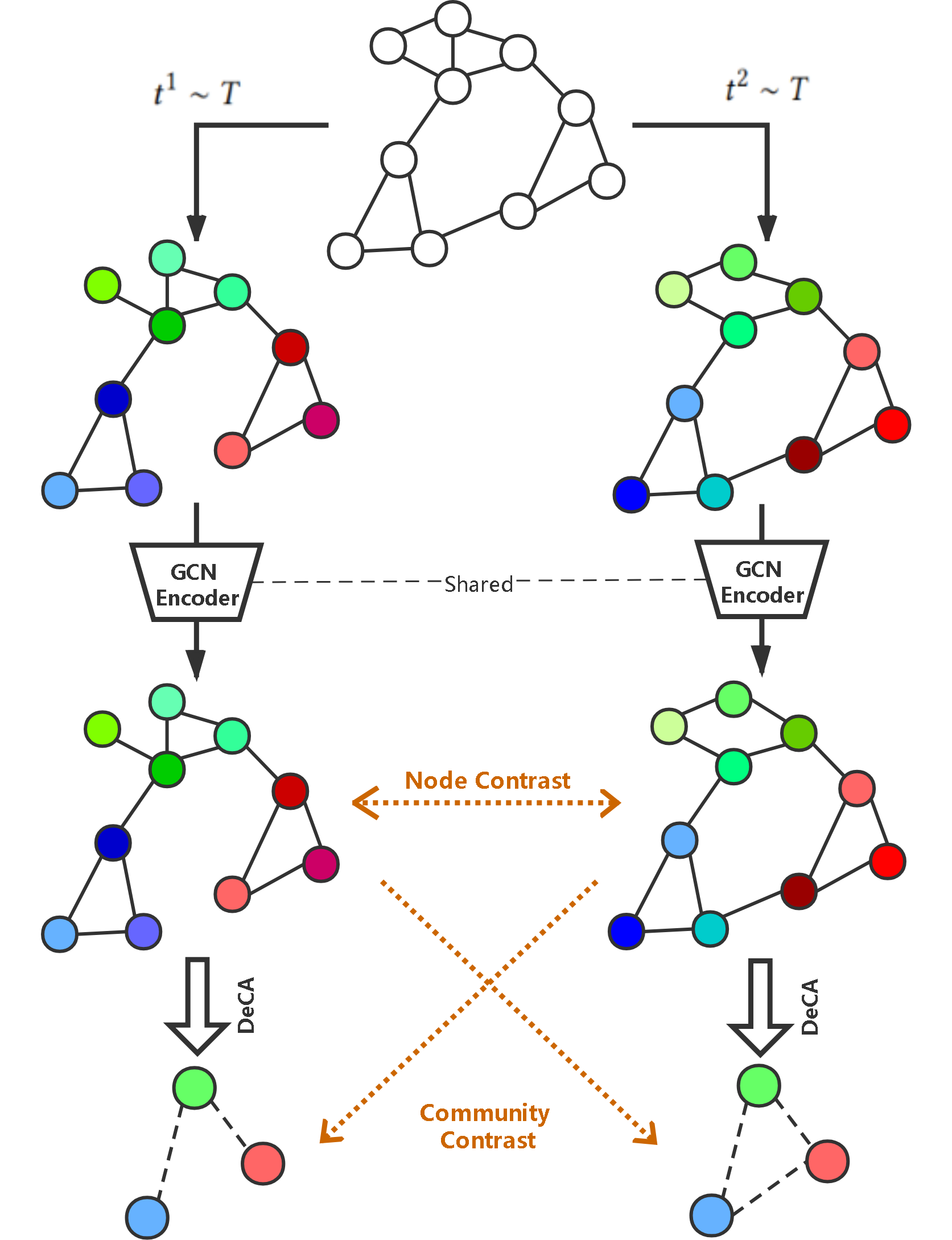}
\caption{Overview of $gCooL$. We firstly generate two views by graph augmentations, secondly encode the node attributes with a shared encoder, and then detect communities with $DeCA$ and learn node representations with $ReSC$.}
\end{figure}

\section{Preliminaries\label{sec:pre}}
In this section, we introduce the similarity measurement used in this paper (Sec.~\ref{subsec:sim}), the basic concepts of contrastive learning (Sec.~\ref{contrast_learn}) and community detection (Sec.~\ref{community_detect}). Then we give a brief definition of attributed multiplex graphs in Sec.~\ref{heter_net}. The notations used in this paper are summarized in Table~\ref{notation}.
\begin{table}[t!]
\renewcommand{\arraystretch}{0.9}%
\caption{Notations.}\label{notation}
\begin{tabular}{c|l}
\hline
Symbol                  & Description                            \\ \hline
$G$                     & attributed graph                       \\
$\mathscr{G}$           & attributed multiplex graph             \\
$E$                     & edge set                               \\
$\widetilde{E}$         & augmented edge set                     \\
$\bm{X}$                & attribute matrix                       \\
$\widetilde{\bm{X}}$    & augmented attribute matrix             \\
$\bm{X}^r$              & attribute matrix of view $r$           \\
$\bm{A}$                & adjacency matrix                        \\
$\widetilde{\bm{A}}$    & extended adjacency matrix               \\
$C$                     & community                              \\
$\bm{\Phi}$             & community centroid matrix              \\
$\bm{R}$                & community assignment matrix            \\
$\bm{F}$                & community density matrix               \\\hline
$E(\cdot)$              & edge count function                    \\
$d(\cdot)$              & edge density function                  \\
$H(\cdot)$              & entropy                                \\
$I(\cdot;\cdot)$        & mutual information                     \\
$\delta_c(\cdot,\cdot)$ & exponent cosine similarity             \\
$\delta_e(\cdot,\cdot)$ & Gaussian RBF similarity                \\
$f_\theta(\cdot)$       & GCN encoder with parameters $\theta$   \\
$g_w(\cdot)$            & linear classifier with weights $w$     \\ \hline
$T$                     & random augmentation                    \\
$D_{intra}$             & intra-community density score, scalar  \\
$D_{inter}$             & inter-community density score, scalar  \\\hline
\end{tabular}
\end{table}

\subsection{Similarity Measurement\label{subsec:sim}}
We adopt two classic similarity functions to determine the similarity between two nodes: \textbf{exponent cosine similarity}~\cite{rahutomo2012semantic}
\begin{equation}
    \delta_c(\bm{x}_1,\bm{x}_2)=exp\left\{\frac{{\bm{x}_1}^T\bm{x}_2/\tau}{||\bm{x}_1||\cdot||\bm{x}_2||}\right\},\label{cos}
\end{equation}
in which we exponentiate the cosine similarity for non-negativity, and \textbf{Gaussian RBF similarity}~\cite{bugmann1998normalized}
\begin{equation}
    \delta_e(\bm{x}_1,\bm{x}_2)=exp\left\{-||\bm{x}_1-\bm{x}_2||^2/\tau^2\right\},\label{eud}
\end{equation}
where $\tau$ is the sensitivity factor.

\subsection{Contrastive Learning\label{contrast_learn}}
Contrastive learning aims at distinguishing similar (positive) and dissimilar (negative) pairs of data points, encouraging the agreement of the similar pairs and the disagreement of the dissimilar ones~\cite{chuang2020debiased}. The general process dating back to \cite{becker1992self} can be described as: For a given data point $\bm{x}$, let $\bm{x}^1$ and $\bm{x}^2$ be differently augmented data points derived from $\bm{x}$. $\bm{z}^1$ and $\bm{z}^2$ are the representations by passing $\bm{x}^1$ and $\bm{x}^2$ through a shared encoder. The mutual information~\cite{cover1991information} (or its variants such as \cite{oord2018representation}) between the two representations is maximized. The motivation of contrastive learning is to learn representations that are invariant to perturbation introduced by the augmentation schemes~\cite{xiao2020should}, and thus are robust to inherent noise in the dataset.

A widely used objective for constrastive learning is InfoNCE~\cite{oord2018representation}, which applies instance-wise contrastive objective by sampling the same instances in different views as positive pairs and all other pairs between different views as negative pairs. The limitation of InfoNCE is that it uniformly samples negative pairs, ignoring the semantic similarities between some instances.

\subsection{Community Detection\label{community_detect}}
Communities in graphs are groups of nodes that are more strongly connected among themselves than others. Community structures widely exist in many real-world graphs, such as sociology, biology and transportation systems~\cite{newman2018networks}. Detecting communities in a graph has become a primary approach to understand how structure relates to high-order information inherent to the graph~\cite{huang2021survey}.

\subsubsection{Formulation}
For a graph $G=(V,E)$, with its node set $V=\{v_1,v_2,......,v_M\}$ and edge set $E=\{(v_{i_1},v_{j_1}),(v_{i_1},v_{j_1}),......,(v_{i_M},v_{j_M})\}$, we define one of its communities $C$ as a sub-graph: $C=(V^C,E^C)$, where $V^C\subseteq V$ and $E^C=E\cap(V^C\times V^C)$.

\subsubsection{Modularity}
A widely used evaluation for community partitions is the modularity~\cite{newman2004finding}, defined as:
\begin{equation}
    m=\frac{1}{2M}\sum_{i,j}\left[\bm{A}[i,j]-\frac{d_id_j}{2M}\right]r(i,j),
\end{equation}
where $\bm{A}$ is the adjacency matrix, $d_i$ is the degree of the $i$-th node, and $r(i,j)$ is $1$ if node $i$ and $j$ are in the same community, otherwise is $0$. The modularity measures the impact of each edge to the local edge density ($d_id_j/2M$ represents the expected local edge density), and is easily disturbed by the changing of edges.

\subsubsection{Other notations\label{subsubsec:other}}
We define the edge count function over the adjacency matrix: 
\begin{equation}
    E(C)=\sum_{i,j}\mathbb{1}\left\{\bm{A}^C[i,j]\neq0\right\},
\end{equation}
where $\bm{A}^C$ is the adjacency matrix for community $C$. The edge density function compares the real edge count to the maximal possible number of edges in the given community $C_k$:
\begin{equation}
    d(k)=\frac{E(C_k)}{|C_k|\left(|C_k|-1\right)}.\label{density}
\end{equation}

\subsection{Attributed Multiplex Graph\label{heter_net}}
\begin{figure}[ht]
    \includegraphics[width=5cm]{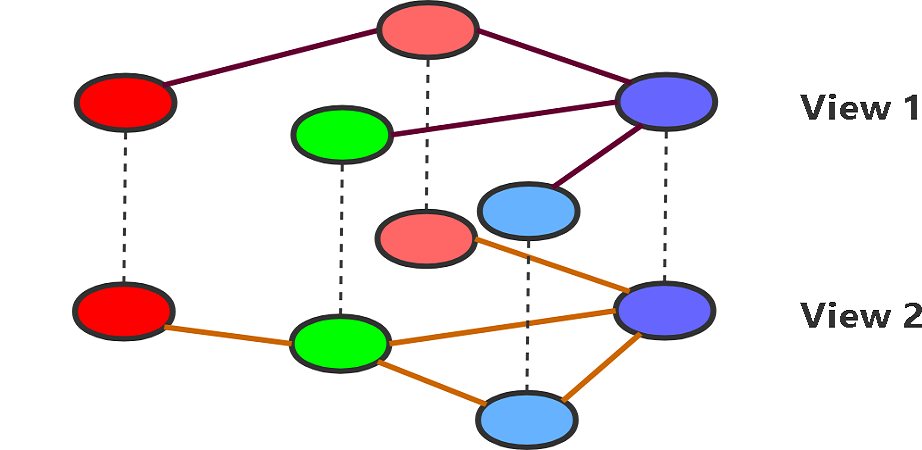}
    \caption{An example of attributed multiplex graph. It consists of two views with shared node attributes and different sets of edges.}
    \label{heterogeneous}
\end{figure}

Multiplex graphs are also known as multi-dimensional graphs~\cite{ma2018multi} or multi-view graphs~\cite{shi2018mvn2vec, fu2020view, jing2021graph}, and are comprised of multiple single-view graphs with shared nodes and attributes but different graph structures (usually with different types of link)~\cite{chang2015heterogeneous}. Formally, an attributed multiplex graph is $\mathscr{G}=\{G^1,G^2,......,G^R\}$, where $R\in\mathbbm{N}_+$ and each $G^r=(V,E^r)$ is an attributed graph. If the number of views $R=1$, $\mathscr{G}=\{G^1\}$ is equivalent to the attributed graph $G^1$. We show an example of attributed multiplex graph in Fig.~\ref{heterogeneous}.

\section{Methodology\label{sec:method}}
In this section, as important components of the proposed $gCooL$ model, we first introduce Dense Community Detection ($DeCA$) algorithm in Sec.~\ref{DeCA}, and then in Sec.~\ref{contrast}, we propose the Reweighted Self-supervised Cross-contrastive ($ReSC$) training scheme. Further discussion on adaption to multiplex graphs is presented in Sec.~\ref{multi}.

\subsection{Dense Community Aggregation\label{DeCA}}
\begin{figure}[ht]
\includegraphics[width=8cm]{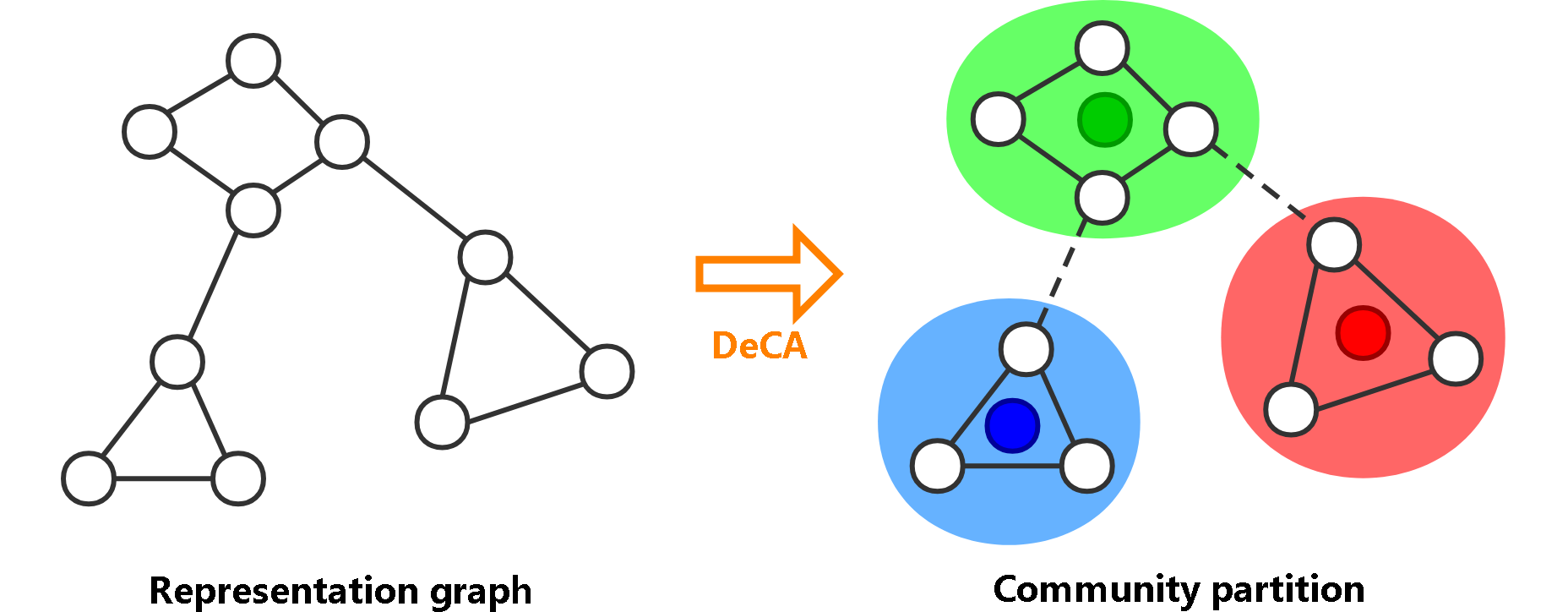}
\caption{Dense Community Aggregation. Each node is softly assigned to community centroids with probabilities, and the community centroids are learned in end-to-end training with Eq.~\ref{LDeCA}.}
\label{fig:DeCA}
\end{figure}

Node-level GCL methods are vulnerable to the problem caused by pairing structurally close nodes as negative samples. For this issue, we propose a novel Dense Community Aggregation ($DeCA$) for detecting structurally related communities of a single graph view. Our method is inspired by the modularity~\cite{newman2004finding} in graphs, which measures the local edge density in communities. However, modularity is easily perturbed by variations of edges~\cite{liu2020deep}, which limits its robustness in detecting communities (with edges dynamically changing in each epochs). Therefore, our goal is to enhance the robustness of modularity, and to further extend the modularity by maximizing edge density in each community, while minimizing the edge density between different communities. The $DeCA$ is carried out by end-to-end training and is illustrated in Fig.~\ref{fig:DeCA}.

We learn the community partition by training a randomly initialized centroid matrix $\bm{\Phi}$ in an end-to-end fashion, where each $\bm{\Phi}[k,:]$ represents the center of the $k$-th community.

First, we assign each node in the graph softly to community centroids with probabilities (computed by similarity function Eq.~\ref{cos} or Eq.~\ref{eud}). Specifically, we define a community assignment matrix $\bm{R}$, where each $\bm{R}[i,:]$ is a normalized similarity vector measuring the distance between the $i$-th node and all community centroids. Formally, $\bm{R}$ is computed by
\begin{equation}
    \bm{R}=normalize(\delta(f_\theta(\bm{X},\bm{A}),\bm{\Phi})),
\end{equation}
where $\delta(\cdot)$ is the similarity function defined in Sec.~\ref{subsec:sim}, $f_\theta(\cdot)$ is the graph encoder with parameters $\theta$, and $normalize(\cdot)$ normalizes the probability of each community by dividing it by the sum of all probabilities and holds $\sum_j\bm{R}[i,j]=1$ for each $i$.

Second, we employ two objectives for training community partition (see Table~\ref{notation} and Sec.~\ref{subsubsec:other} for symbol definitions): \textbf{intra-community density}
\begin{equation}
    D_{intra}=\frac{1}{N}\sum_{i,j}\sum_{k}\left[\bm{A}[i,j]-d(k)\right]\bm{R}[i,k]\bm{R}[j,k],\label{DL}
\end{equation}
and \textbf{inter-community density}
\begin{equation}
    D_{inter}=\frac{1}{N(N-1)}\sum_{i,j}\sum_{k_1\neq k_2}\bm{A}[i,j]\bm{R}[i,k_1]\bm{R}[j,k_2].\label{DW}
\end{equation}
These two objectives measure the impact of each edge on the community edge density (rather than the local edge density used in modularity~\cite{newman2004finding}). Specifically, in Eq.~\ref{DL} and Eq.~\ref{DW}, the term $\bm{A}[i,j]-d(k)$ and $\bm{A}[i,j]-0$ represent the gap between real local density ($\bm{A}[i,j]$) and the expected density ($d(k)$ for intra-community and $0$ for inter-community). The centroid matrix $\bm{\Phi}$ will be updated to reach a reasonable community partition, by minimizing the joint objective:
\begin{equation}
    l_{DeCA}(\bm{R})=\lambda_wD_{inter}-D_{intra},\label{lDeCA}
\end{equation}
where $\lambda_w$ is the co-efficient. Moreover, for computational efficiency, we slightly modify the form of $l_{DeCA}$ in our actual implementation (shown in Appendix~\ref{ap:para}). Finally, we combine the $l_{DeCA}$ objectives for two graph views, and carry out dense community aggregation simultaneously for them:
\begin{equation}
    L_{DeCA}=\frac{1}{2}\left[l_{DeCA}(\bm{R^1})+l_{DeCA}(\bm{R^2})\right].\label{LDeCA}
\end{equation}

\subsection{Reweighted Self-supervised Cross-contrastive Training\label{contrast}}
We propose the Reweighted Self-supervised Cross-contrastive ($ReSC$) training scheme in this section. We firstly apply graph augmentations to generate two graph views and secondly apply \textbf{node contrast} and \textbf{community contrast} simultaneously to consider both node-level and community-level information. We introduce node-community pairs as additional negative samples to address the problem of pairing nodes in the same communities as negative samples.

\subsubsection{Graph augmentation}
We define the graph augmentation $T$ as randomly masking attributes and edges with a specific probability~\cite{zhu2021graph}. First, we define the attribute masks as random noise vector $\bm{m}$, where each dimension of it is independently drawn from a Bernoulli distribution: $\bm{m}[i]\sim Bernoulli(1-p_v)$. The augmented attribute matrix is computed by
\begin{equation}
    \widetilde{\bm{X}}=\left[\bm{X}[1,:]\odot\bm{m};\bm{X}[2,:]\odot\bm{m};......;\bm{X}[N,:]\odot\bm{m}\right]',
\end{equation}
where $\odot$ is the Hadamard product~\cite{horn1990hadamard} and $\bm{X}$ is the original attribute matrix. Second, we generate the augmented edge set $\widetilde{E}$ by randomly dropping edges from the original edge set $E$~\cite{zhu2020deep} with probability
\begin{equation}
    P\left\{(v_1,v_2)\in\widetilde{E}\right\}=1-p_e, \forall(v_1,v_2)\in E.
\end{equation}

We apply the above graph augmentations (denoted as $t^1,t^2\sim T$ respectively for 2 independent augmentations) to generate two graph views: $(\bm{X}^1,\bm{A}^1)=t^1(\bm{X},\bm{A})$ and $(\bm{X}^2,\bm{A}^2)=t^2(\bm{X},\bm{A})$ for $t^1,t^2\sim T$. Then, we obtain their representations by a shared GCN encoder: $\bm{Z}^1=f_\theta(\bm{X}^1,\bm{A}^1)$ and $\bm{Z}^2=f_\theta(\bm{X}^2,\bm{A}^2)$.

\subsubsection{Node contrast}
After generating two graph views, we employ node contrast and community contrast simultaneously to learn the node representations. We introduce a contrastive loss based on the InfoNCE~\cite{oord2018representation} to contrast node-node pairs:
\begin{equation}
    I_{NCE}(\bm{Z}^1;\bm{Z}^2)=-log\sum_{i}\frac{\delta(\bm{Z}^1[i,:],\bm{Z}^2[i,:])}{\sum_{j}\delta(\bm{Z}^1[i,:],\bm{Z}^2[j,:])}.
\end{equation}
We apply the node contrast symmetrically for the two graph views:
\begin{equation}
    L_{node}=\frac{1}{2}\left[I_{NCE}(\bm{Z}^1;\bm{Z}^2)+I_{NCE}(\bm{Z}^2;\bm{Z}^1)\right],\label{Lnode}
\end{equation}
which distinguishes negative pairs in the two views, and enforces maximizing the consistency between positive pairs~\cite{lin2021prototypical}.

\subsubsection{Community contrast}
The community contrast is based on the result of $DeCA$ in Sec.~\ref{DeCA}. First, we obtain the community centers by training the randomly initialized community centroids matrix $\bm{\Phi}$ with Eq.~\ref{LDeCA}. Second, we adopt a re-weighted cross-contrastive objective to contrast node representations of one view with the community centroids of the other view (a cross-contrastive fashion). We are inspired by the cross-prediction scheme in \cite{caron2020unsupervised} and introduce the cross-contrast into InfoNCE objective~\cite{oord2018representation} to maximize the community consistency between two views, enforcing that the node representations in one community are pulled close to those of the counterpart community in the other view. Formally, the community contrast is computed by
\begin{equation}
\begin{aligned}
&l_{com}(\bm{Z},\bm{\Phi})= \\
&-log\sum_{i}\frac{\delta(\bm{Z}[i,:],\bm{\Phi}[k_i,:])}{\delta(\bm{Z}[i,:],\bm{\Phi}[k_i,:])+\sum_{k_i\neq k}w(i,k)\cdot\delta(\bm{Z}[i,:],\bm{\Phi}[k,:])},
\end{aligned}
\end{equation}
where the $i$-th node is assigned to the $k_i$-th community. Here, $w(i,k)=exp\left\{-\gamma||\bm{Z}[i,:]-\bm{\Phi}[k,:]||^2\right\}$ is the RBF weight function (slightly different from Gaussian RBF similarity in Eq.~\ref{eud}), which gives more attention to the more similar community pairs. In this objective, the similarity of node representations within the same communities are maximized since they are contrasted positively to the same centroids, while in different communities, node representations are separated by negative contrast. Similarly, we compute the contrastive objective symmetrically for the two generated graph views:
\begin{equation}
    L_{com}=\frac{1}{2}\left[l_{com}(\bm{Z}^1,\bm{\Phi}^2)+l_{com}(\bm{Z}^2,\bm{\Phi}^1)\right],\label{Lmeta}
\end{equation}
where $\bm{Z}^1$ and $\bm{\Phi}^2$ are the node representation matrices of view 1 and view 2 respectively.

\subsubsection{Joint objective}
We propose the $\alpha$-decay coefficient to combine $L_{node}$, $L_{DeCA}$ and $L_{com}$ into a joint objective:
\begin{equation}
    L=L_{node}+\alpha(t) L_{DeCA}+[1-\alpha(t)]L_{com},\label{joint}
\end{equation}
where the coefficient $\alpha(t)=exp\left\{-t/\eta\right\}$ would decay smoothly with the proceeding of training ($t$ refers to the epoch). We observe that the community partition would be stabilized within a few hundreds of epochs by $DeCA$ training (see Sec.~\ref{visual}), while the training the $gCooL$ model usually takes thousands of epochs. To this end, we apply the $\alpha$-decay to focus mainly on training community partition at first, and gradually divert the focus to learning node representations.

In summary, the $ReSC$ training process is shown in Algorithm~\ref{alo1}.
\begin{algorithm}
    \caption{The $ReSC$ training.}
    \label{alo1}
    Inputs: attribute matrix $\bm{X}$, adjacency matrix $\bm{A}$; \\
    Output: graph encoder $f_\theta(\cdot)$; \\
    \For {t = 1,2,......}{
        Sample two random graph augmentations $t^1,t^2\sim T$; \\
        Generate two graph views, $(\bm{X}^1,\bm{A}^1)=t^1(\bm{X},\bm{A})$ and $(\bm{X}^2,\bm{A}^2)=t^2(\bm{X},\bm{A})$; \\
        Encode the two views by a shared GCN encoder, $\bm{Z}^1=f_\theta(\bm{X}^1,\bm{A}^1)$ and $\bm{Z}^2=f_\theta(\bm{X}^2,\bm{A}^2)$; \\
        Obtain $L_{node}$, $L_{DeCA}$ and $L_{com}$ to compute the joint loss: $L=L_{node}+\alpha(t) L_{DeCA}+[1-\alpha(t)]L_{com}$; \\
        Update parameters $\theta$ by stochastic gradient descent on joint objective $L$.
    }
\end{algorithm}

\subsection{Adaptation to Multiplex graphs\label{multi}}
The proposed $gCooL$ framework can be naturally adapted to multiplex graphs with a few modifications on the training and inference process. We apply contrastive training between different graph views, which consider the dependence between them.
\subsubsection{Training}
In multiplex graphs, we no longer need to generate graph views through graph augmentation, since the different views in a multiplex graph are naturally multi-viewed data. We propose to detect communities ($DeCA$) and learn node representations ($ReSC$) on each pair of views. The modified training process is shown in Algorithm~\ref{heter_train}.
\begin{algorithm}
    \caption{The $ReSC$ training on multiplex graphs.}
    \label{heter_train}
    Inputs: attribute matrix $\bm{X}$, adjacency matrices $\bm{A}^1$...$\bm{A}^R$; \\
    Output: graph encoder $f_\theta(\cdot)$; \\
    \For {t = 1,2,......}{
        Encode the R graph views by a shared GCN encoder, \\ $\bm{Z}^1=f_\theta(\bm{X}^1,\bm{A}^1)$, \\ $\bm{Z}^2=f_\theta(\bm{X}^2,\bm{A}^2)$, \\
        ...... \\
        $\bm{Z}^R=f_\theta(\bm{X}^R,\bm{A}^R)$; \\
        Obtain $L_{node}$, $L_{DeCA}$ and $L_{com}$ to compute the joint loss for each pair of views $(G^i,G^j)$ such that $i\neq j$; \\
        Update parameters $\theta$ by stochastic gradient descent on joint objective $L$ in Eq.~\ref{joint} (adding up objectives of each pair of views).
    }
\end{algorithm}
\subsubsection{Inference}
At the inference time, we propose to combine classification results of each view by classifier fusion (a post-ensemble fashion): Given the results of $R$ independent classifiers, we label the final prediction according to the confidence of each classifier (i.e. the maximal value of the output softmax distribution~\cite{hendrycks2016baseline}). We choose the result with the strongest confidence as the final prediction.

\section{Experiments\label{sec:exp}}
In this section, we evaluate the proposed $gCooL$ model with respect to diverse down-stream tasks on multiple real graphs. We introduce the specifications of experiments in Sec.~\ref{exp_setup}, and provide detailed quantitative and visual results in Sec.~\ref{quant} and Sec.~\ref{visual} respectively.

\subsection{Experimental Setup\label{exp_setup}}
\subsubsection{Datasets}
We use 6 real graphs, including Amazon-Computers, Amazon-Photo, Coauthor-CS, WikiCS, IMDB and DBLP, to evaluate the performance on node classification and node clustering. The detailed statistics of all datasets are summarized in Table~\ref{data_sta}.
\begin{itemize}
    \item \textbf{Amazon-Computers} and \textbf{Amazon-Photo}~\cite{shchur2018pitfalls} are two graphs of co-purchase relations, where nodes are products and there exists an edge between two products when they are bought together. Each node has a sparse bag-of-words attribute vector encoding product reviews.
    \item \textbf{Coauthor-CS}~\cite{shchur2018pitfalls} is an academic network based on the Microsoft Academic Graph. The nodes are authors and edges are co-author relations. Each node has a sparse bag-of-words attribute vector encoding paper keywords of the author.
    \item \textbf{WikiCS}~\cite{mernyei2020wiki} is a reference network from Wikipedia. The nodes are articles and edges are hyperlinks between the articles. Node attributes are computed as the average of GloVe~\cite{pennington2014glove} word embedding of articles.
    \item \textbf{IMDB}~\cite{wang2019heterogeneous} is a movie network with two types of links: co-actor and co-director. The attribute vector of each node is a bag-of-words feature of movie plots. The nodes are labeled with Action, Comedy or Drama.
    \item \textbf{DBLP}~\cite{wang2019heterogeneous} is a paper graph with three types of links: co-author, co-paper and co-term. The attribute vector of each node is a bag-of-words feature vector of paper abstracts.
\end{itemize}
\begin{table}[ht]\small%
\setlength\tabcolsep{1.8pt}%
\renewcommand{\arraystretch}{0.8}%
\caption{Dataset statistics.}
\label{data_sta}
\begin{tabular}{c|ccccc}
\hline
Dataset               & Type        & Edges  & Nodes               & Attributes          & Classes          \\ \hline
Amazon-Computers\footnotemark[1]       & co-purchase & 245,778   & 13,381                 & 767                   & 10                 \\
Amazon-Photo\footnotemark[1]          & co-purchase & 119,043   & 7,487                  & 745                   & 8                  \\
Coauthor-CS\footnotemark[1]           & co-author   & 81,894    & 18,333                 & 6,805                  & 15                 \\
WikiCS\footnotemark[2]                & reference   & 216,123   & 11,701                 & 300                   & 10                 \\ \hline
\multirow{2}{*}{IMDB\footnotemark[3]} & co-actor    & 66,428    & \multirow{2}{*}{3,550} & \multirow{2}{*}{2,000} & \multirow{2}{*}{3} \\
                      & co-director & 13,788    &                       &                       &                    \\ \hline
\multirow{3}{*}{DBLP\footnotemark[4]} & co-author   & 144,738   & \multirow{3}{*}{7,907} & \multirow{3}{*}{2,000} & \multirow{3}{*}{4} \\
                      & co-paper    & 90,145    &                       &                       &                    \\
                      & co-term     & 57,137,515 &                       &                       &                    \\ \hline
\end{tabular}
\end{table}
\footnotetext[1]{\url{https://github.com/shchur/gnn-benchmark/raw/master/data/npz/}}
\footnotetext[2]{\url{https://github.com/pmernyei/wiki-cs-dataset/raw/master/dataset}}
\footnotetext[3]{\url{https://www.imdb.com/}}
\footnotetext[4]{\url{https://dblp.uni-trier.de/}}

\subsubsection{Evaluation protocol}
We follow the evaluations in \cite{zhu2021graph}, where each graph encoder is trained in a self-supervised way. Then, the encoded representations are used to train a classifier for node classification, and fit a K-means model for comparing baselines on node clustering. We train each graph encoder for five runs with randomly selected data splits (for WikiCS, we use the given 20 data splits) and report the average performance with the standard deviation.

For the node classification task, we measure the performance with Micro-F1 and Macro-F1 scores. For the node clustering task, we measure the performance with Normalized Mutual Information (NMI) score: $NMI=2I(\bm{\hat{Y}};\bm{Y})/[H(\bm{\hat{Y}})+H(\bm{Y})]$, where $\bm{\hat{Y}}$ and $\bm{Y}$ refer to the predicted cluster indexes and class labels respectively, and Adjusted Rand Index (ARI): $ARI=RI-\mathbbm{E}[RI]/(max\{RI\}-\mathbbm{E}[RI])$, where $RI$ is the Rand Index~\cite{rand1971objective}, which measures the similarity between two cluster indexes and class labels.

\subsubsection{Baselines}
We compare the $gCooL$ methods ($gCooL_c$ for exponent cosine similarity and $gCooL_e$ for Gaussian RBF similarity) with two sets of baselines:
\begin{itemize}
    \item \textbf{On single-view graphs}, we consider baselines as 1) traditional methods including node2vec~\cite{grover2016node2vec} and DeepWalk~\cite{perozzi2014deepwalk}, 2) supervised methods including GCN~\cite{kipf2016semi}, and 3) unsupervised methods including MVGRL~\cite{hassani2020contrastive}, DGI~\cite{velivckovic2018deep}, HDI~\cite{jing2021hdmi}, graph autoencoders (GAE and VGAE)~\cite{kipf2016variational} and GCA~\cite{zhu2021graph}.
    \item \textbf{On multiplex graphs}, we consider baselines as 1) methods with single-view representations including node2vec~\cite{grover2016node2vec}, DeepWalk~\cite{perozzi2014deepwalk}, GCN~\cite{kipf2016semi} and DGI~\cite{velivckovic2018deep}, and 2) methods with multi-view representations including CMNA~\cite{chu2019cross}, MNE~\cite{yang2019deep}, HAN~\cite{wang2019heterogeneous}, DMGI~\cite{park2020unsupervised} and HDMI~\cite{jing2021hdmi}.
\end{itemize}
Additionally, we compare different clustering baselines including K-means, Spectral biclustering (SBC)~\cite{kluger2003spectral} and modularity~\cite{newman2004finding} to show the effectiveness of our proposed $DeCA$ ($DeCA_c$ for exponent cosine similarity and $DeCA_e$ for Gaussian RBF similarity).

\subsubsection{Implementation}
We use a 2-layer GCN~\cite{kipf2016semi} as the graph encoder for each deep learning baseline. We use Adam optimizer~\cite{kingma2015adam} to optimize each model. Before applying the $ReSC$ objective, we pass the representations of two graph views through a shard projection module (a 2-layer MLP) with the same input, hidden-layer and output dimensions. The number of communities is set according to the number of classes in each graph. More implementation details are listed in Appendix~\ref{hyper}.

\subsection{Quantitative Results\label{quant}}
\subsubsection{Overall performance}
We present the quantitative results of node classification on single-view graphs (Table~\ref{tab:class}), node clustering on single-view graphs (Table~\ref{tab:cluster}), and node classification on multiplex graphs (Table~\ref{tab:class_heter}). $gCooL$ outperforms the state-of-the-art methods on multiple tasks, and even outperforms the supervised methods.
% The raw data of experimental results is listed in Appendix~\ref{app:quant}.
\begin{table*}[ht]
\renewcommand{\arraystretch}{0.8}%
\caption{Overall performance on node classification (in percentage).}
\label{tab:class}
\begin{tabular}{c|cc|cc|cc|cc}
\hline
Dataset     & \multicolumn{2}{c|}{Amazon-Computers}      & \multicolumn{2}{c|}{Amazon-Photo}         & \multicolumn{2}{c|}{Coauthor-CS}          & \multicolumn{2}{c}{WikiCS}                \\ \hline
Metric      & Micro-F1            & Macro-F1            & Micro-F1            & Macro-F1            & Micro-F1            & Macro-F1            & Micro-F1            & Macro-F1            \\ \hline
RawFeatures & 73.82±0.01          & 70.10±0.04          & 78.45±0.04          & 76.10±0.01          & 90.40±0.02          & 89.01±0.06          & 72.00±0.03          & 70.28±0.09          \\
Node2vec    & 84.38±0.08          & 82.65±0.08          & 89.72±0.08          & 87.39±0.07          & 85.11±0.06          & 82.93±0.11          & 71.84±0.09          & 70.44±0.03          \\
DeepWalk    & 85.63±0.09          & 84.02±0.10          & 89.36±0.10          & 86.92±0.02          & 84.71±0.23          & 82.63±0.19          & 74.25±0.06          & 72.68±0.15          \\
GCN         & 86.43±0.56          & 83.99±0.61          & 92.51±0.23          & 90.47±0.32          & 93.04±0.28          & 91.02±0.38          & 77.11±0.08          & 75.61±0.19          \\
MVGRL       & 87.42±0.07          & 85.92±0.11          & 91.74±0.09          & 89.93±0.09          & 92.11±0.10          & 90.50±0.12          & 77.50±0.08          & 75.62±0.00          \\
DGI         & 83.88±0.50          & 79.30±0.42          & 91.60±0.24          & 89.31±0.16          & 92.08±0.68          & 90.78±0.68          & 75.35±0.17          & 73.74±0.20          \\
HDI         & 85.43±0.13          & 80.74±0.25          & 90.09±0.10          & 88.70±0.16          & 89.98±0.14          & 86.73±0.17          & 75.72±0.55          & 68.05±0.80          \\
GAE         & 85.18±0.21          & 83.33±0.17          & 91.68±0.14          & 89.66±0.09          & 90.00±0.75          & 88.31±0.68          & 70.17±0.05          & 68.27±0.05          \\
VGAE        & 86.44±0.25          & 83,72±0.12          & 92.24±0.08          & 90.04±0.17          & 92.08±0.08          & 90.11±0.06          & 75.56±0.20          & 74.12±0.10          \\
GCA         & 87.67±0.49          & 85.88±0.30          & 92.39±0.21          & 91.05±0.13          & 92.87±0.03          & 90.76±0.01          & 78.24±0.01          & 74.47±0.02          \\ \hline
$gCooL_c$    & \textbf{88.85±0.14} & 87.42±0.28          & \textbf{93.18±0.12} & \textbf{92.05±0.17} & \textbf{93.32±0.02} & \textbf{91.65±0.03} & \textbf{78.74±0.04} & \textbf{75.92±0.06} \\
$gCooL_e$    & 88.74±0.09          & \textbf{87.53±0.26} & 92.79±0.17          & 91.57±0.29          & 93.31±0.01          & 91.63±0.03          & \textbf{78.74±0.03} & 75.88±0.02          \\ \hline
\end{tabular}
\end{table*}

\begin{table*}[ht]
\renewcommand{\arraystretch}{0.8}%
\caption{Overall performance on node clustering.}
\label{tab:cluster}
\begin{tabular}{c|cc|cc|cc|cc}
\hline
Dataset  & \multicolumn{2}{c|}{Amazon-Computers}        & \multicolumn{2}{c|}{Amazon-Photo}           & \multicolumn{2}{c|}{Coauthor-CS}            & \multicolumn{2}{c}{WikiCS}                  \\ \hline
Metric   & NMI                  & ARI                  & NMI                  & ARI                  & NMI                  & ARI                  & NMI                  & ARI                  \\ \hline
MVGRL    & 0.244±0.000          & 0.141±0.001          & 0.344±0.040          & 0.239±0.039          & 0.740±0.010          & 0.627±0.009          & 0.263±0.010          & 0.102±0.011          \\
DGI      & 0.318±0.020          & 0.165±0.020          & 0.376±0.030          & 0.264±0.030          & 0.747±0.010          & \textbf{0.629±0.011} & 0.310±0.020          & 0.131±0.018          \\
HDI      & 0.347±0.011          & 0.216±0.006          & 0.429±0.014          & 0.307±0.011          & 0.726±0.008          & 0.607±0.016          & 0.238±0.002          & 0.105±0.000          \\
GAE      & 0.441±0.000          & 0.258±0.000          & 0.616±0.010          & 0.494±0.008          & 0.731±0.010          & 0.614±0.010          & 0.243±0.020          & 0.095±0.018          \\
VGAE     & 0.423±0.000          & 0.238±0.001          & 0.530±0.040          & 0.373±0.041          & 0.733±0.000          & 0.618±0.001          & 0.261±0.010          & 0.082±0.008          \\
GCA      & 0.426±0.001          & 0.246±0.001          & 0.614±0.000          & 0.494±0.000          & 0.735±0.008          & 0.618±0.010          & 0.299±0.002          & 0121±0.003           \\ \hline
$gCooL_c$ & 0.452±0.000          & \textbf{0.284±0.000} & 0.619±0.000          & 0.508±0.000          & 0.750±0.021          & 0.624±0.083          & 0.318±0.011          & \textbf{0.145±0.024} \\
$gCooL_e$ & \textbf{0.474±0.018} & 0.277±0.024          & \textbf{0.632±0.000} & \textbf{0.524±0.001} & \textbf{0.753±0.005} & 0.621±0.007          & \textbf{0.322±0.009} & 0.140±0.010          \\ \hline
\end{tabular}
\end{table*}

\begin{table}[ht]
\renewcommand{\arraystretch}{0.8}%
\caption{Overall classification performance on multiplex graphs.}
\label{tab:class_heter}
\begin{tabular}{c|cc|cc}
\hline
Dataset  & \multicolumn{2}{c|}{IMDB}        & \multicolumn{2}{c}{DBLP}       \\ \hline
Metric   & Micro-F1       & Macro-F1       & Micro-F1       & Macro-F1       \\ \hline
node2vec & 0.550          & 0.533          & 0.547          & 0.543          \\
DeepWalk & 0.550          & 0.532          & 0.537          & 0.533          \\
GCN      & 0.611          & 0.603          & 0.717          & 0.734          \\
DGI      & 0.606          & 0.598          & 0.720          & 0.723          \\
CMNA     & 0.566          & 0.549          & 0.561          & 0.566          \\
MNE      & 0.574          & 0.552          & 0.562          & 0.566          \\
HAN      & 0.607          & 0.599          & 0.708          & 0.716          \\
DMGI     & 0.648          & 0.648          & 0.766          & 0.771          \\
HDMI     & 0.658          & 0.650          & 0.811          & 0.820          \\ \hline
$gCooL_c$ & \textbf{0.672} & \textbf{0.670} & \textbf{0.832} & \textbf{0.840} \\
$gCooL_e$ & 0.671          & 0.668          & \textbf{0.832} & 0.839          \\ \hline
\end{tabular}
\end{table}

Moreover, the proposed $DeCA$ (used to cluster nodes based on its community partition) outperforms traditional clustering methods and the former modularity~\cite{newman2004finding} on encoded representations (shown in Table~\ref{tab:cluster1}), which illustrates that $DeCA$ is more suitable for deep-learning-based clustering and addresses the instability of modularity.
\begin{table*}[ht]
\renewcommand{\arraystretch}{0.8}%
\caption{Performance on node clustering, comparing different clustering methods on randomly initialized GCN encoder.}
\label{tab:cluster1}
\begin{tabular}{c|cc|cc|cc|cc}
\hline
Dataset    & \multicolumn{2}{c|}{Amazon-Computers}       & \multicolumn{2}{c|}{Amazon-Photo}           & \multicolumn{2}{c|}{Coauthor-CS}            & \multicolumn{2}{c}{WikiCS}                  \\ \hline
Metric     & MNI                  & ARI                  & MNI                  & ARI                  & MNI                  & ARI                  & MNI                  & ARI                  \\ \hline
K-means    & 0.192±0.019          & 0.086±0.013          & 0.225±0.012          & 0.121±0.006          & 0.497±0.004          & 0.315±0.008          & 0.047±0.010          & 0.022±0.002          \\
SBC        & 0.160±0.000          & 0.088±0.000          & 0.100±0.000          & 0.020±0.000          & 0.415±0.000          & 0.261±0.000          & 0.039±0.000          & 0.024±0.000          \\
modularity & 0.204±0.036          & 0.118±0.038          & 0.233±0.027          & 0.124±0.032          & 0.447±0.017          & 0.280±0.022          & 0.139±0.063          & 0.083±0.052          \\
$DeCA_c$   & 0.213±0.009          & 0.114±0.018          & 0.237±0.024          & 0.126±0.032          & 0.476±0.015          & 0.305±0.020          & 0.140±0.060          & 0.080±0.055          \\
$DeCA_e$   & \textbf{0.363±0.009} & \textbf{0.262±0.010} & \textbf{0.384±0.022} & \textbf{0.273±0.025} & \textbf{0.523±0.016} & \textbf{0.358±0.018} & \textbf{0.210±0.041} & \textbf{0.136±0.060} \\ \hline
\end{tabular}
\end{table*}

\subsubsection{Ablation study}
We compare the different combinations of $L_{node}$ (Eq.~\ref{Lnode}), $L_{DeCA}$ (Eq.~\ref{LDeCA}) and $L_{com}$ (Eq.~\ref{Lmeta}) on WikiCS in terms of node classification and node clustering. The results are listed in Table~\ref{tab:abl}. We find that, first, applying node contrast is crucial to the performance (given that the combinations with $L_{node}$ outperform others) since it makes the representations more distinguishable, second, community contrast is beneficial to the overall performance (given that $L_{node}+L_{com}$ outperforms $L_{node}$) since it considers community information in contrastive training, and third, the performance of community contrast is boosted significantly when it is used with $DeCA$ (comparing $L_{com}$ with $L_{DeCA}+L_{com}$). Moreover, we also show the visual ablation study in Table~\ref{tab:tsne}.
\begin{table*}[ht]
\renewcommand{\arraystretch}{0.6}%
\caption{Ablation study on node classification and node clustering, comparing combinations of objectives on WikiCS.}
\label{tab:abl}
\begin{tabular}{ccc|cc|cccc}
\hline
$L_{node}$   & $L_{DeCA}$   & $L_{com}$    & $\delta_c(\cdot)$ & $\delta_e(\cdot)$ & Micro-F1(\%)        & Macro-F1(\%)        & NMI                  & ARI                  \\ \hline
$\checkmark$ &              &              & $\checkmark$      &                   & 75.09±0.29          & 69.83±0.24          & 0.254±0.005          & 0.120±0.007          \\
$\checkmark$ &              &              &                   & $\checkmark$      & 75.71±0.58          & 68.52±0.55          & 0.228±0.015          & 0.127±0.030          \\
             &              & $\checkmark$ & $\checkmark$      &                   & 65.78±0.29          & 57.25±0.36          & 0.191±0.001          & 0.102±0.001          \\
             &              & $\checkmark$ &                   & $\checkmark$      & 63.48±0.67          & 53.78±0.90          & 0.188±0.007          & 0.085±0.005          \\
$\checkmark$ &              & $\checkmark$ & $\checkmark$      &                   & 77.29±0.20          & 74.19±0.23          & 0.273±0.012          & 0.125±0.020          \\
$\checkmark$ &              & $\checkmark$ &                   & $\checkmark$      & \textbf{77.47±0.28} & \textbf{74.43±0.30} & \textbf{0.274±0.003} & \textbf{0.139±0.001} \\
             & $\checkmark$ & $\checkmark$ & $\checkmark$      &                   & 72.88±0.21          & 69.97±0.24          & 0.222±0.020          & 0.120±0.030          \\
             & $\checkmark$ & $\checkmark$ &                   & $\checkmark$      & 68.18±0.14          & 64.88±0.16          & 0.227±0.012          & 0.123±0.034          \\ \hline\hline
$\checkmark$ & $\checkmark$ & $\checkmark$ & $\checkmark$      & \textbf{}         & \textbf{78.74±0.04} & \textbf{75.92±0.06} & 0.318±0.011          & \textbf{0.145±0.024} \\
$\checkmark$ & $\checkmark$ & $\checkmark$ &                   & $\checkmark$      & \textbf{78.74±0.03} & 75.88±0.02          & \textbf{0.322±0.009} & 0.140±0.010          \\ \hline
\end{tabular}
\end{table*}

\subsection{Visual Evaluations\label{visual}}
We illustrate the significance of $DeCA$ by visualizing the \textbf{edge density} and \textbf{class entropy} of the assigned communities. We evaluate each checkpoint for five times and show the means and deviations. We compare the results with traditional clustering methods (K-means and Spectral biclustering~\cite{kluger2003spectral}) and the former modularity~\cite{newman2004finding}. We also visualize the node representations for ablation study.
% The raw data for the visualization is listed in Appendix~\ref{app:visual}.

\subsubsection{Edge density}
\begin{figure}[ht]
\centering
\includegraphics[width=8cm,height=5cm]{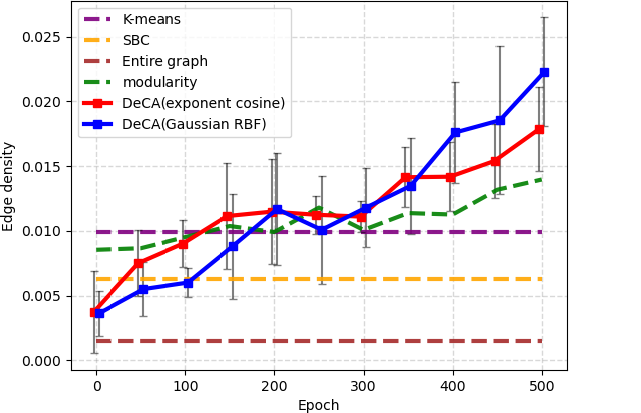}
\caption{Visualization of edge density comparison.}
\label{fig:edge}
\end{figure}

The edge density is based on Eq.~\ref{density} and computed by the average density of all communities:
\begin{equation}
    ED=\frac{1}{K}\sum_{k=1}^Kd(k).
\end{equation}
It is used to measure how $DeCA$ learns the community partition that maximizes the intra-community density (see Sec.~\ref{DeCA}). Fig.~\ref{fig:edge} shows that, after a few hundreds of epochs, $DeCA$ stably outperforms other clustering methods.

\subsubsection{Class entropy}
\begin{figure}[ht]
\centering
\includegraphics[width=8cm,height=5cm]{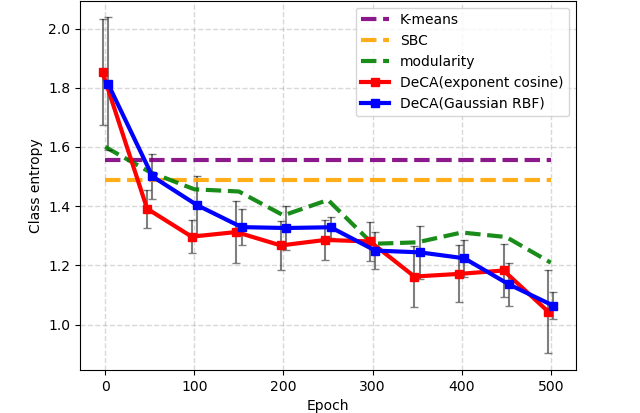}
\caption{Visualization of class entropy comparison.}
\label{fig:class}
\end{figure}

The class entropy is a measure of the homogeneity of class labels in a community (the extent to which a community contains one major class, or has low entropy). We argue that a good community partition should distinguish structurally separated nodes, which is, in other words, distinguishing nodes of different classes. The class entropy is computed as the average entropy of class labels in all communities:
\begin{equation}
    CH=-\frac{1}{K}\sum_{k=1}^K\sum_{c}P_k(c)logP_k(c),
\end{equation}
where $P_k(c)$ is the occurrence frequency of class $c$ in the $k$-th community. Fig.~\ref{fig:class} shows that, after a few hundreds of epochs, $DeCA$ stably outperforms other clustering methods.

\subsubsection{Visualization of node representations.}
To see how the node representations are distributed, we use t-SNE~\cite{van2008visualizing} to reduce the dimension of node representations for visualization. The node representations of each class are distributed more separately when applying $L_{DeCA}$ as well as $L_{com}$, which illustrates the effectiveness of our proposed method. The results are shown in Table~\ref{tab:tsne}.
\begin{table*}[t]
\caption{Ablation study on WikiCS by visualizing node representations with t-SNE.}
\label{tab:tsne}
\begin{tabular}{c|c|c|c|c|c}
\hline
& $L_{node}$                                                                                                                                    & $L_{com}$                                                                                                                                     & $L_{node}+L_{com}$                                                                                & $L_{DeCA}+L_{com}$                                                                                & Full                                                                                                                                           \\ \hline
\rotatebox{90}{exponent cosine}                                             & \begin{minipage}[b]{0.3\columnwidth}\centering\raisebox{-.05\height}{\includegraphics[width=\linewidth]{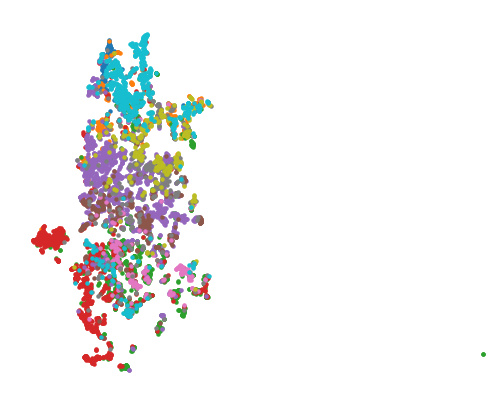}}\end{minipage} & \begin{minipage}[b]{0.3\columnwidth}\centering\raisebox{-.05\height}{\includegraphics[width=\linewidth]{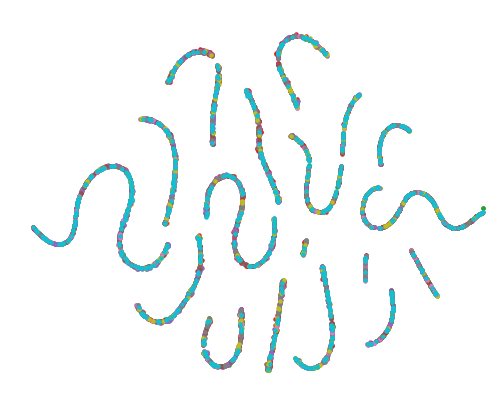}}\end{minipage} & \begin{minipage}[b]{0.3\columnwidth}\centering\raisebox{-.05\height}{\includegraphics[width=\linewidth]{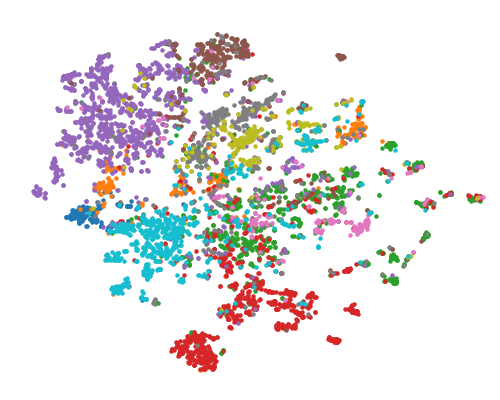}}\end{minipage} & \begin{minipage}[b]{0.3\columnwidth}\centering\raisebox{-.05\height}{\includegraphics[width=\linewidth]{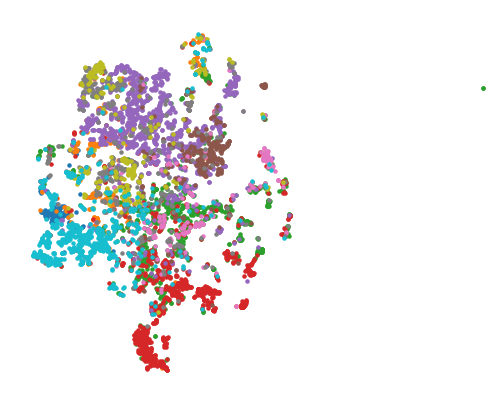}}\end{minipage} & \begin{minipage}[b]{0.3\columnwidth}\centering\raisebox{-.05\height}{\includegraphics[width=\linewidth]{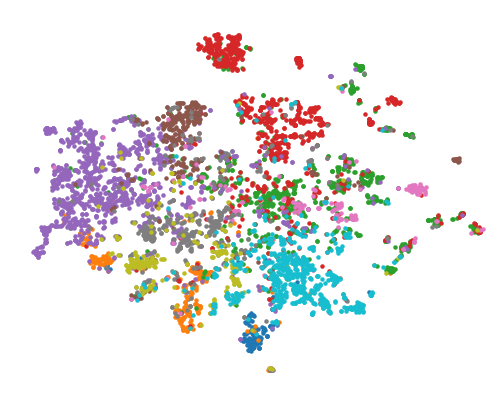}}\end{minipage}  \\ \hline
\rotatebox{90}{Gaussian RBF}                                             & \begin{minipage}[b]{0.3\columnwidth}\centering\raisebox{-.13\height}{\includegraphics[width=\linewidth]{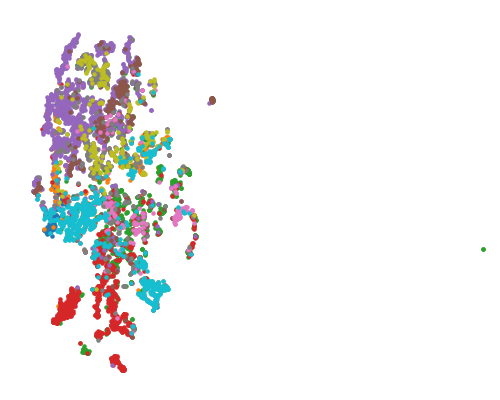}}\end{minipage} & \begin{minipage}[b]{0.3\columnwidth}\centering\raisebox{-.13\height}{\includegraphics[width=\linewidth]{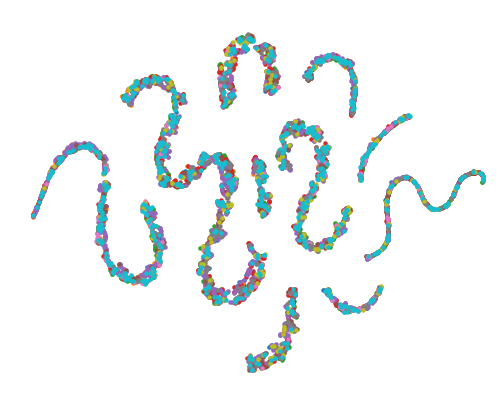}}\end{minipage} & \begin{minipage}[b]{0.3\columnwidth}\centering\raisebox{-.13\height}{\includegraphics[width=\linewidth]{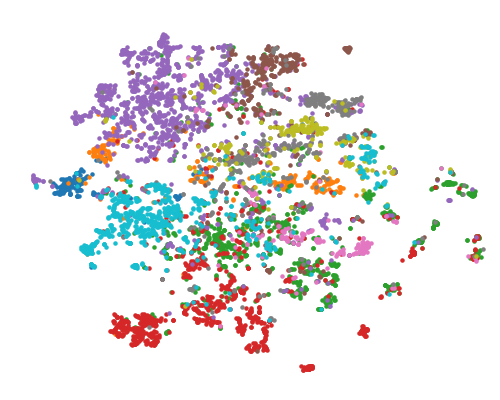}}\end{minipage} & \begin{minipage}[b]{0.3\columnwidth}\centering\raisebox{-.13\height}{\includegraphics[width=\linewidth]{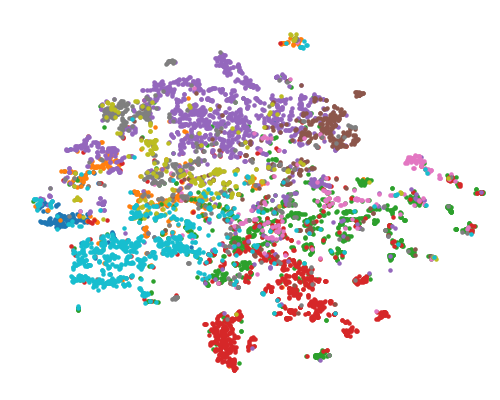}}\end{minipage} & \begin{minipage}[b]{0.3\columnwidth}\centering\raisebox{-.13\height}{\includegraphics[width=\linewidth]{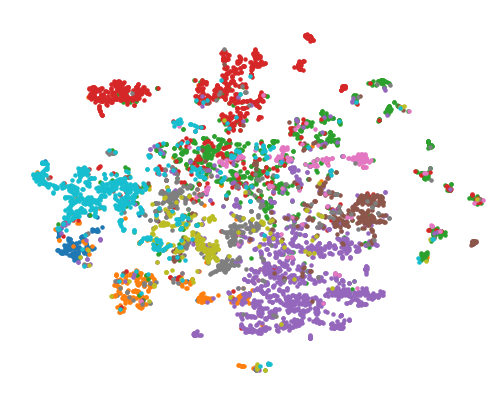}}\end{minipage} \\ \hline
\end{tabular}
\end{table*}

\section{Related Works\label{sec:relate}}
In this section, we briefly review the related works, including graph contrastive learning (Sec.~\ref{sec:gcl}) and community detection (Sec.~\ref{sec:cd}).

\subsection{Graph Contrastive Learning\label{sec:gcl}}
Graph contrastive learning~\cite{li2019graph} proposed to pull similar nodes into close positions in the representation space while pushing dissimilar nodes apart. \cite{liu2021graph,wu2021self} comprehensively review the development of GCL. GRACE~\cite{zhu2020deep} proposed a simplified framework to maximize the agreement between two graph views. GCA~\cite{zhu2021graph} proposed an adaptive augmentation algorithm to generate graph views. GCC~\cite{qiu2020gcc} designed a pre-training task to learn structural graph representations. HeCo~\cite{wang2021self} employed network schema and meta-path views to employ contrastive training. DGI~\cite{velivckovic2018deep}, GMI~\cite{peng2020graph} and HDI~\cite{jing2021hdmi} randomly shuffle node attributes to generate negative samples. MVGRL~\cite{hassani2020contrastive} employed graph diffusion to transform graphs.

For multiplex graphs, DGMI~\cite{park2020unsupervised} and HDMI~\cite{jing2021hdmi} sample negative pairs by randomly shuffling the attributes. MNE~\cite{zhang2018scalable}, mvn2vec~\cite{shi2018mvn2vec} and GATNE~\cite{cen2019representation} use random walk to sample negative pairs.

\subsection{Community Detection\label{sec:cd}}
Community detection is a primary approach to understand how structure relates to high-order information inherent to the graph. \cite{huang2021survey} reviewed the recent progress in community detection. CommDGI~\cite{zhang2020commdgi} employed modularity for community detection. \cite{li2020optimization} introduced the identifiability of communities. RWM~\cite{luo2020local} used random walk to detect local communities. 
Destine \cite{xu2021destine,xu2021dense} detects dense subgraphs on multi-layered networks.
% \cite{xin2017deep} uses CNN model to detect communities in topologically incomplete graphs.

\section{Conclusion\label{sec:con}}
In this paper, we propose a novel Graph Communal Contrastive Learning ($gCooL$) framework to learn the community partition of structurally related communities by a Dense Community Aggregation ($DeCA$) algorithm, and to learn the graph representation with a reweighted self-supervised cross-contrastive ($ReSC$) training scheme considering the community structures. The proposed $gCooL$ consistently achieves the state-of-the-art performance on multiple tasks and can be naturally adapted to multiplex graphs. We show that community information is beneficial to the overall performance in graph representation learning.

\begin{acks}
B. Jing and H. Tong are partially supported by National Science Foundation under grant No. 1947135, the NSF Program on Fairness in AI in collaboration with Amazon under award No. 1939725, NIFA 2020-67021-32799, and Army Research Office (W911NF2110088). The content of the information in this document does not necessarily reflect the position or the policy of the Government or Amazon, and no official endorsement should be inferred. The U.S. Government is authorized to reproduce and distribute reprints for Government purposes notwithstanding any copyright notation here on.
\end{acks}

\bibliographystyle{ACM-Reference-Format}
\bibliography{main}

\appendix
\newpage
~~~
\newpage
\section{Implementation Details\label{hyper}}
We conduct all of our experiments with PyTorch Geometric 1.7.2~\cite{fey2019fast} and PyTorch 1.8.1~\cite{paszke2019pytorch} on a NVIDIA Tesla V100 GPU (with 32 GB available video memory). All single-viewed graph datasets used in experiments are available in PyTorch Geometric libraries\footnote[5]{\url{https://pytorch-geometric.readthedocs.io/en/latest/modules/datasets.html}}, and all multiplex graph datasets are available in \footnote[6]{\url{https://www.dropbox.com/s/48oe7shjq0ih151/data.tar.gz?dl=0}}. Our hyperparameter specifications are listed in Table~\ref{tab:hyper}. Those hyperparameters are determined empirically by gird search based on the settings of \cite{zhu2021graph}.
\begin{table}[ht]\footnotesize%
\setlength\tabcolsep{1.8pt}
\caption{Hyperparameter specifications.}
\label{tab:hyper}
\begin{tabular}{c|cccc|cc}
\hline
Hyperparameter           & \begin{tabular}[c]{@{}c@{}}Amazon-\\ Computers\end{tabular} & \begin{tabular}[c]{@{}c@{}}Amazon\\ -Photo\end{tabular} & \begin{tabular}[c]{@{}c@{}}Coauthor\\ -CS\end{tabular} & WikiCS & IMDB   & DBLP  \\ \hline
Training epochs          & 2900                                                        & \begin{tabular}[c]{@{}c@{}}2900\\ /1400\end{tabular}    & 600                                                    & 1000   & 50     & 50    \\
Hidden-layer dimension   & 2558                                                        & 256                                                     & 7105                                                   & 1368   & 8096   & 4256  \\
Representation dimension & 512                                                         & 128                                                     & 300                                                    & 384    & 2048   & 128   \\
Learning rate            & 0.01                                                        & 0.1                                                     & 0.0005                                                 & 0.01   & 0.0001 & 0.001 \\
Activation function      & RReLU                                                       & ReLU                                                    & RReLU                                                  & PReLU  & ReLU   & RReLU \\
$\eta$                   & 1000                                                        & 500                                                     & 500                                                    & 500    & 50     & 50    \\
$\tau$                   & 0.2                                                         & 0.3                                                     & 0.4                                                    & 0.4    & 1.0    & 1.0   \\
$\lambda_w$              & 0.9                                                         & 0.1                                                     & 0.9                                                    & 1.0    & 0.5    & 0.5   \\
$\gamma$                 & 2e-5                                                        & 8e-5                                                    & 8e-5                                                   & 8e-5   & 2e-4   & 2e-4  \\
$p_v$                    & 0.25                                                        & 0.1                                                     & 0.3                                                    & 0.1    & 0.1    & 0.1   \\
$p_e$                    & 0.45                                                        & 0.4                                                     & 0.2                                                    & 0.2    & 0.0    & 0.0   \\ \hline
\end{tabular}
\end{table}

\section{Number of communities}
\begin{figure}[ht]
\centering
\includegraphics[width=8cm]{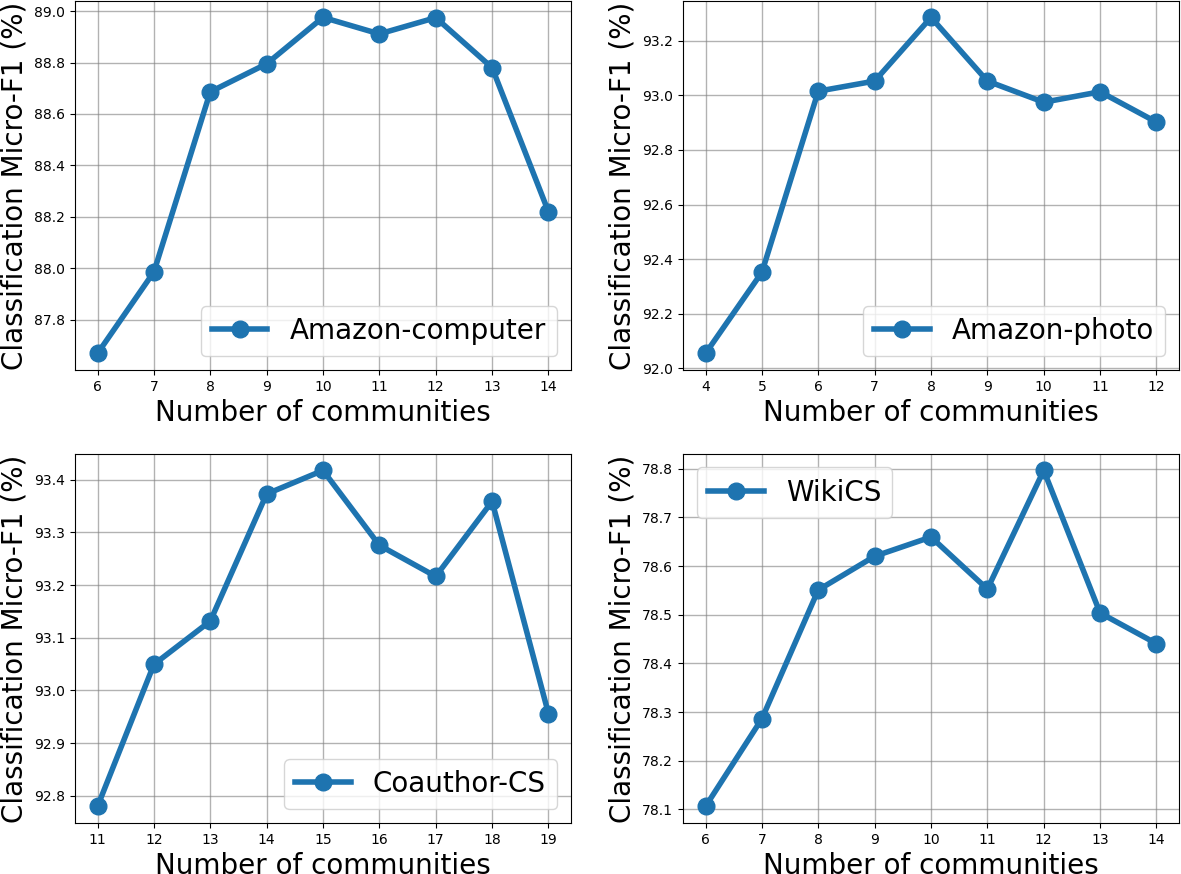}
\caption{Comparison of community number on node classification.}
\label{fig:com_number}
\end{figure}

We compare varying numbers of communities in the vicinity of the numbers of classes on each single-viewed graph dataset (refer to Fig.~\ref{fig:com_number}). Although the performances remain stable for nearly all settings, we recommend to set the number of communities according to the actual number of classes.

\section{Parallelization of Dense Community Aggregation\label{ap:para}}
In $DeCA$, The intra-community density objective is computationally costly and hard to vectorize since the expected edge density is dependent on the community choice. To address this problem, we derive a lower bound of $D_{intra}$:
\begin{equation}
\begin{aligned}
D_{intra} &\geq \frac{1}{N}\sum_{i,j}\sum_{k}\left[\bm{A}[i,j]-\max_{\kappa}\{d(\kappa)\}\right]\bm{R}[i,k]\bm{R}[j,k] \\
    &=\frac{1}{N}\sum_{i,j}\sum_{k}\widetilde{\bm{A}}[i,j]\bm{R}[i,k]\bm{R}[j,k] \\
    &=\widetilde{D}_{intra},
\end{aligned}
\end{equation}
where $\widetilde{\bm{A}}=\bm{A}-\max_\kappa d(\kappa)\bm{I}$ is the extended adjacency matrix, and $\widetilde{D}_{intra}=\inf D_{intra}$ is the lower bound. We use $\widetilde{D}_{intra}$ to replace $D_{intra}$ in Eq.~\ref{lDeCA}.

Next, we employ the community density matrix $\bm{F}=\bm{R}'\bm{A}\bm{R}$ and $\widetilde{\bm{F}}=\bm{R}'\widetilde{\bm{A}}\bm{R}$ to vectorize the objective $D_{inter}$ and $\widetilde{D}_{intra}$. The entry of $\bm{F}$ is $\bm{F}[u,v]=\sum_{i}\bm{R}[i,u]\cdot(\bm{A}\bm{R})[i,v]=\sum_{i,j}\bm{A}[i,j]\bm{R}[i,u]\bm{R}[j,v]$, which naturally accords with the form of Eq.~\ref{DL} and Eq.~\ref{DW}. Therefore, these two objectives can be re-formalized as
\begin{equation}
    \widetilde{D}_{intra}=\frac{1}{N}tr(\widetilde{\bm{F}}),
\end{equation}
and
\begin{equation}
    D_{inter}=\frac{1}{N(N-1)}\left[\sum_{i,j}\bm{F}[i,j]-tr(\bm{F})\right].
\end{equation}

Finally, the vectorized $DeCA$ objective is
\begin{equation}
\begin{aligned}
l_{DeCA}(\bm{R})&=\lambda_wD_{inter}-\widetilde{D}_{intra} \\
    &=\frac{\lambda_w}{N(N-1)}\left[\sum_{i,j}\bm{F}[i,j]-tr(\bm{F})\right]-\frac{1}{N}tr(\widetilde{\bm{F}}),
\end{aligned}
\end{equation}
which can be directly computed by the CUDA operators~\cite{paszke2019pytorch}.

\end{document}